\documentclass[letterpaper,12pt]{article}
\usepackage{amsmath, amsthm, amssymb, fancyhdr, fancybox, graphicx, url, color, wasysym}
\usepackage{natbib}
\usepackage{sectsty}
\sectionfont{\large}
\usepackage{bbm,pbox}
\usepackage[export]{adjustbox}
\usepackage{array,float}

\usepackage[margin=0.75in]{geometry}
\usepackage{times}

\usepackage{graphicx}

\begin{document}

\title{Assessing the impacts of mutations to the structure of COVID-19 spike protein via sequential Monte Carlo}

\author{Samuel W.K. Wong\footnote{ Address for correspondence: Department of Statistics and Actuarial Science, University of Waterloo, Waterloo, ON, Canada.  E-mail: samuel.wong@uwaterloo.ca}
	\\ Department of Statistics and Actuarial Science, University of Waterloo
}

\date{June 8, 2020}

\maketitle{}

\begin{abstract}	
Proteins play a key role in facilitating the infectiousness of the 2019 novel coronavirus.  A specific spike protein enables this virus to bind to human cells, and a thorough understanding of its 3-dimensional structure is therefore critical for developing effective therapeutic interventions.  However, its structure may continue to evolve over time as a result of mutations.  In this paper, we use a data science perspective to study the potential structural impacts due to ongoing mutations in its amino acid sequence.  To do so, we identify a key segment of the protein and apply a sequential Monte Carlo sampling method to detect possible changes to the space of low-energy conformations for different amino acid sequences.  Such computational approaches can further our understanding of this protein structure and complement laboratory efforts.
\end{abstract}

\section{Introduction}

The COVID-19 disease is an ongoing public health concern since the 2019 novel coronavirus, formally known as SARS-CoV-2 \citep{gorbalenya2020species}, has spread throughout the world.  Much research has been devoted to understand the mechanism by which the virus attacks human cells \citep[e.g.,][]{hoffmann2020sars,zhang2020angiotensin}.  Such a scientific understanding is critical to the ongoing development of therapeutic interventions and vaccines for COVID-19, see, e.g., \citet{amanat2020sars} for a review of efforts to date.  Proteins play key roles in facilitating the entry of viruses into cells, and specifically for SARS-CoV-2, the spike (S) glycoprotein that protrudes from its viral membrane \citep{walls2020structure}.  Genome sequencing of SARS-CoV-2 has shown that this new coronavirus has moderate genetic similarity to the SARS-associated coronavirus (formally SARS-CoV) that caused the 2002-2003 outbreak, having between 75-80\% of their genetic material in common \citep{lu2020genomic}.  In particular, both SARS-CoV and SARS-CoV-2 utilize a spike protein to bind to the ACE2 (Angiotensin Converting Enzyme 2) receptor in human cells \citep{ou2020characterization}.  Laboratory work has now shown that the 3-dimensional (3-D) structures of these two spike proteins are broadly similar; however, they are sufficiently different such that antibodies for SARS-CoV are not effective against SARS-CoV-2 \citep{wrapp2020cryo}.

As a result of Anfinsen's Nobel prize-winning work in the 1970s, it has been known that a protein generally has a stable 3-D structure that is largely determined by its amino acid sequence \citep{anfinesen1973principles}.  A broad consequence of this discovery, that has been verified in the subsequent decades of laboratory experimental work, is that similarities in amino acid sequences often correspond to similarities in 3-D structure \citep{krissinel2007relationship}.  However, laboratory techniques for determining protein structure are laborious and cannot be applied to all possible amino acid sequences of interest, e.g., in drug design applications \citep{khoury2014protein}.  Further, the atomic coordinates of amino acids sometimes cannot be ascertained, and will thus be missing data in the published 3-D structures \citep{brandt2008seqatoms}.  Thus, to complement laboratory work, there has been much interest in developing computational methods to tackle the protein folding problem, that is, to predict the 3-D structure of a protein based on its amino acid sequence \citep{friesner2002computational}.  While much progress has been made in the past five decades towards improving the accuracy of such computation-based structure predictions \citep{dill2012protein}, the protein folding problem is still considered unsolved in general.  The most successful competitor to date, as assessed by the most recent biannual blinded protein folding competition in 2018 known as CASP13 (Critical Assessment of Structure Prediction)\footnote{http://predictioncenter.org}, was the deep learning method AlphaFold developed by Google's DeepMind team \citep{alquraishi2019alphafold}.

We briefly summarize the timeline of key milestones in the scientific understanding of the SARS-CoV-2 spike protein, to provide context for the contribution of this paper.  Using modern RNA sequencing technology, scientists were quickly able to determine the original first known SARS-CoV-2 genome \citep{wu2020new}, including the amino acid sequence of its S protein, by early January 2020.  On the basis of this sequence and its similarities with related proteins, including previously-determined 3-D structures of SARS-CoV and MERS, preliminary 3-D structure predictions of the SARS-CoV-2 S protein could be generated using computational methods\footnote{e.g., https://www.ipd.uw.edu/2020/02/rosettas-role-in-fighting-coronavirus/}.   Subsequently, thanks to worldwide attention and rapid research efforts, a UT Austin laboratory was the first to release a 3-D structure of the SARS-CoV-2 S protein in mid-February 2020, using cyro-EM techniques \citep{wrapp2020cryo}.  A limitation was that several segments of the protein were not successfully determined, and thus missing from the published structure.  Nonetheless, this ground-breaking work verified that it is a key segment of the S protein, known as the receptor-binding domain (RBD), that has the ability to bind with human ACE2 when the RBD is in its ``open'' state.  Further, \citet{wrapp2020cryo} showed that in comparison with the corresponding RBD of the SARS-CoV S protein, the two have noticeable similarity in their overall structures, but they have significant deviations in some local segments.  In particular, their binding sites are sufficiently different such that SARS-CoV antibodies cannot recognize SARS-CoV-2.  By early March 2020, the 3-D structure of the RBD of the SARS-CoV-2 S protein, in a structural complex bound together with ACE2, was also published via laboratory efforts \citep{yan2020structural}.

These 3-D structures are publicly available in the Protein Data Bank (PDB) \citep{bernstein1977protein}, and are a vital piece of the puzzle in the ongoing development of interventions to neutralize the SARS-CoV-2 virus.  It is also well-known, however, that viral genomes mutate over time with varying rates \citep{drake1993rates}.  Mutations underlie, for example, why the prevalent strain of influenza may be difficult to predict when designing the annual flu vaccine \citep{cohen2017flu}.  Mutations are changes in the viral genome that occur as the virus replicates; possible results of mutations include changes in the amino acid sequence of its proteins -- via additions, deletions, or substitutions of individual amino acids \citep{sanjuan2010viral}.  Thus,  scientists continue to pay close attention to variations in sequenced SARS-CoV-2 genomes over time \citep{tang2020origin}.  At the time of this writing, among sequenced SARS-CoV-2 genomes there have already been two amino acid substitutions observed in a key segment of its S protein RBD, as we shall subsequently detail.  Therefore, there is concern about the impact of such mutations on the protein structure, and in turn on the binding ability and infectiousness of SARS-CoV-2 as it continues to evolve worldwide \citep{jia2020analysis}.  It is infeasible to use laboratory-based structure determination to study all potential sequence mutations, thus we may instead use computational methods to assess the potential structural impacts and complement laboratory efforts.

These considerations provide the main motivation of this paper.  Our goal is to adopt a data science perspective in studying the potential effects of mutations on the 3-D structure of the RBD.  Towards this objective, our specific contributions may be summarized as three major points.  First, by taking the SARS-CoV-2 S protein as a case study, we highlight the importance of the protein structure prediction problem, and introduce the challenges and opportunities it presents to data scientists.  Second, we describe how sequential Monte Carlo sampling methods can be adapted to yield insight into the structural impact of protein sequence mutations.  Third, we discuss our findings based on applying the methodology to a key segment of the SARS-CoV-2 S protein RBD.

\section{The SARS-CoV-2 S protein as a case study:  background on protein sequence and structure}

\subsection{The SARS-CoV-2 S protein sequence}

\begin{figure}[!htbp]
	\centering
	\includegraphics[scale=0.6]{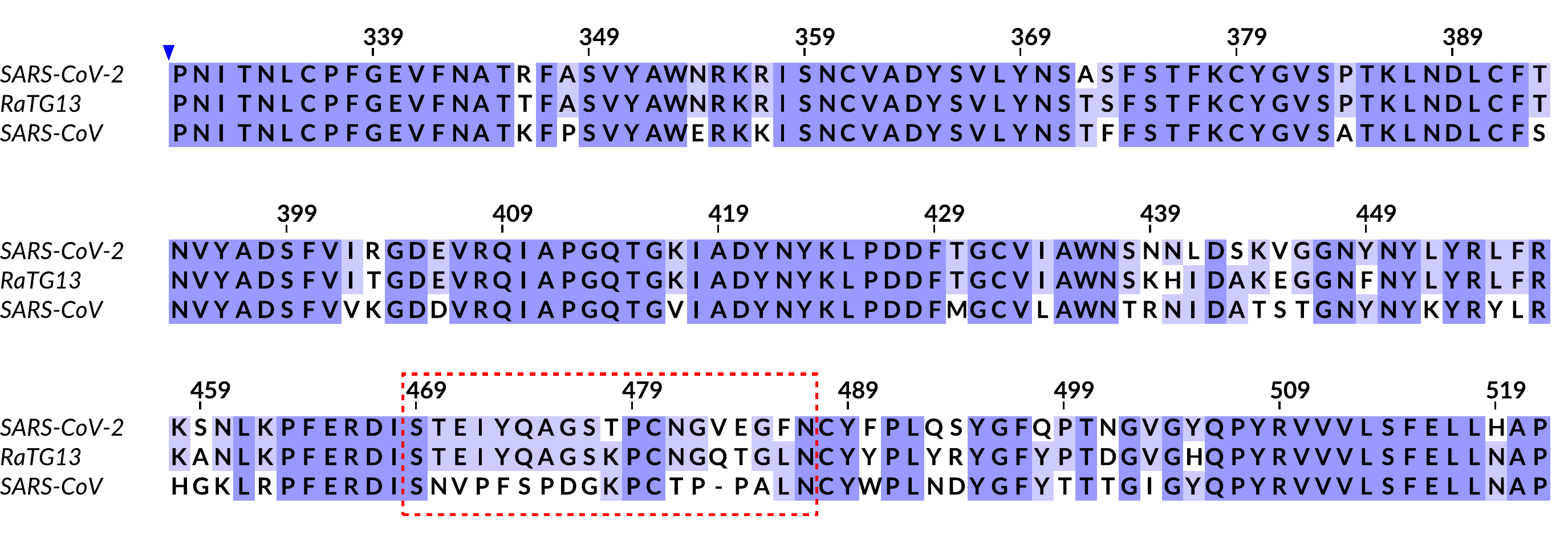}
	\caption{Aligned amino acid sequences for the RBDs of SARS-CoV-2, RaTG13, and SARS-CoV spike glycoproteins. Positions where all three sequences have the same amino acid type are shaded in dark blue, while positions where two of the sequences have the same amino acid type are shaded in light blue.  The red box marks the segment consisting of positions 469 to 487 that has substantive amino acid differences among the three sequences. The position (483) where SARS-CoV has one less amino acid compared to SARS-CoV-2 and RaTG13 is marked by the dash symbol.  Visualized with Jalview software.}
	\label{fig:align}
\end{figure}

The SARS-CoV-2 S protein consists of a linear sequence of 1273 amino acids, and our focus is its RBD which consists of the amino acid positions 329 to 522 \citep{wrapp2020cryo}, and we shall use this SARS-CoV-2 sequence to reference positions for the remainder of the paper.  This sequence may be compared to that the SARS-CoV S protein, which is overall slightly shorter at 1255 amino acids long, with its corresponding RBD located from amino acids 316 to 508.  The two RBDs share 144 (74\%) of their amino acids in  common, and the SARS-CoV-2 RBD has one extra amino acid (Valine) inserted at position 483.  It is also worth noting that there are existing known protein sequences with higher degree of similarity to SARS-CoV-2 than SARS-CoV: the S protein of the bat coronavirus known as RaTG13 \citep{zhou2020pneumonia} has an amino acid sequence that matches 97\% of the SARS-CoV-2 S protein.  Thus RaTG13 S may be even more structurally similar to SARS-CoV-2 S; however, there is no 3-D laboratory-determined structure currently available for RaTG13 S in the PDB.  To visualize, Figure \ref{fig:align} lays out the RBDs of these three sequences in an optimally aligned fashion that shows their matching, or \textit{conserved}, amino acids using the blue shaded colors. There are 20 amino acid types, with each being represented by its one-letter abbreviation.  The dash symbol (--) indicates the position where SARS-CoV has one fewer amino acid compared to SARS-CoV-2 and RaTG13.  At the sequence level, it can be seen that SARS-CoV-2 S and SARS-CoV S differ the most in the segments 437 to 461 and 469 to 487, the latter being indicated by the red box.

\subsection{From sequence to structure}\label{sec:seq2struct}

A standard metric for comparing two 3-D protein structures is the root-mean-square deviation (RMSD) between the corresponding pairs of atomic Cartesian coordinates, after applying the optimal translation and rotation to superimpose the structures.  Specifically, it is standard practice to use \textit{backbone} atoms for RMSD calculations, which are the N, C, C$_\alpha$, and O atoms that are common to all 20 amino acid types and link adjacent amino acids together.  The \textit{side chains} are functional groups extending from the C$_\alpha$ atom and are unique for each of the 20 amino acid types.  That is, we may equivalently say that an amino acid substitution implies that a change in the side chain has occurred at that position.

Based on the SARS-CoV-2 S protein sequence and previously known structures in the PDB, researchers at University of Washington used their Robetta structure prediction server \citep{kim2004protein} to build a preliminary 3-D structure prediction for the entire SARS-CoV-2 S protein\footnote{Available for download from http://new.robetta.org/results.php?id=15652}.  Subsequently, 3-D structures were laboratory-determined and also became publicly available in the PDB:  the accession codes are 6VSB for the full standalone SARS-CoV-2 S protein \citep{wrapp2020cryo}, and 6LZG for the structural complex of the RBD bound together with the ACE2 receptor \citep{yan2020structural}.  Thus this provides a good test case to comment on the accuracy of Robetta's structure prediction, as it represents the efforts of one of the current leading research groups in computational protein folding.

We find that the 6VSB PDB structure (chain A, which has the RBD observed in its ``open''  state that enables binding with human ACE2) is missing the coordinates of all the amino acids in the segment 444 to 491 and two other short RBD segments, likely due to experimental difficulty. In contrast, the 6LZG PDB structure (chain B is the SARS-CoV-2 RBD) has a continuous and complete set of RBD atomic coordinates.  The Robetta RBD structure prediction has an RMSD of 0.933 to the available atomic coordinates of 6VSB, and an RMSD of 2.411 to 6LZG.  However, if the segment 444 to 491 in 6LZG is ignored in the calculation, the RMSD drops to 1.208.  Thus, the Robetta prediction is quite accurate for most of the RBD structure; however, it has difficulty with the segment that includes the positions 469 to 487 noted in Figure \ref{fig:rbdrmsd}.  This result highlights a limitation of current computational approaches:  it is still challenging to build accurate predictions for segments that have little sequence similarity with existing proteins in the PDB.  Therefore, further advances in computational methods for structure prediction continue to be needed.  We address this point in greater detail in Section \ref{sec:bench}, by evaluating and comparing the prediction accuracy achieved by state-of-the-art methods on this challenging segment of the SARS-CoV-2 RBD and the corresponding segment of the SARS-CoV RBD.

\subsection{Structure comparison between SARS-CoV and SARS-CoV-2}

We now specifically examine the available 3-D structures for the RBDs of SARS-CoV-2 and SARS-CoV, using public data from the PDB.  For this purpose we also download the structural complex of the RBD bound together with the ACE2 receptor for SARS-CoV (PDB code: 2AJF) to compare with that of SARS-CoV-2 (PDB code: 6LZG).  These structural complexes have more complete atomic coordinates for the RBD, compared to structures of the full standalone S protein which are missing key RBD segments (PDB codes: 5X58 for SARS-CoV, 6VSB for SARS-CoV-2).  The RMSD may be computed on the two RBDs after alignment; positions 389 to 394 of SARS-CoV-2 are excluded from the calculation as the atomic coordinates atoms of the corresponding positions in the SARS-CoV 3-D structure are missing, and position 483 is also excluded as SARS-CoV-2 has an extra amino acid at that position.  The RMSD between the two RBDs calculated in this way is 1.69, indicating a strong level of similarity in the overall structures.

\begin{figure}[!htbp]
	\centering
	\includegraphics[scale=0.6]{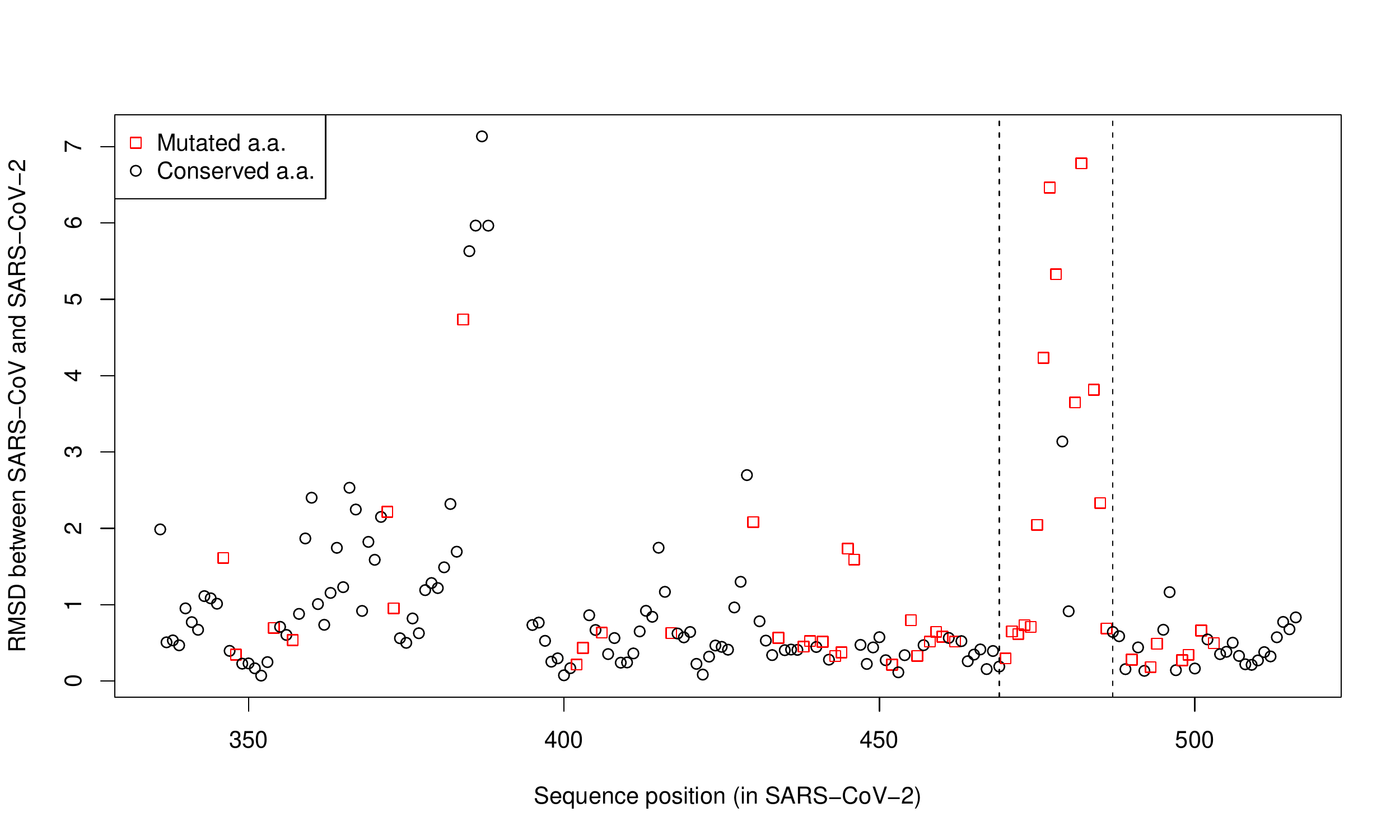}
	\caption{RMSDs between the 3-D structures of the SARS-CoV-2 and SARS-CoV RBDs in their bound complex with ACE2, calculated at each amino acid position.  Black circles indicate positions where the sequences are conserved (i.e., having the same amino acid type), while red squares indicate positions with different amino acid types.  The dashed lines indicate positions 469 and 487, the key segment of interest in this paper.}
	\label{fig:rbdrmsd}
\end{figure}

To examine how this variation is distributed over the length of the RBD, in Figure \ref{fig:rbdrmsd} we plot the RMSDs of the backbone atoms for each amino acid position separately.  Positions where the two protein sequences are conserved are indicated by the circles, while the square markers indicate positions where the amino acid types differ.  From the plot, it can be seen that there are two distinct segments with high RMSD values between the two RBDs:  the first is the five amino acids before the gap of missing coordinates in the SARS-CoV structure, and the second is the segment 469 to 487 as indicated by the dashed lines, where we previously noted a low level of sequence similarity. It can also be observed that while there are many sequence differences in the segment 437 to 461 as well, these have less of an effect on structural difference.  In \citet{yan2020structural}, the authors also note that while there is overall similarity between the two RBDs, there are non-trivial differences in the positions where the RBDs forms chemical bonds with ACE2. In particular, positions 474 and 486 of the SARS-CoV-2 RBD were identified as binding sites, where there is substantial structural dissimilarity with the SARS-CoV RBD.

Since this structural comparison reveals that 469 to 487 is likely a critical segment affected by amino acid substitutions, we will focus in this paper on the potential impact of mutations to this segment.  Indeed, this is also a segment where the SARS-CoV-2 sequence is noticeably distinct from its closest known related sequence RaTG13, with 4 of the 22 substitutions in the RBD occurring this segment; resulting structural differences here may partially explain its inability to infect humans.  Using BLAST \citep{altschul1997gapped} on the \texttt{nr} sequence database on April 23, 2020, we find that there are already two amino acid substitutions observed within the segment 469 to 487 among existing SARS-CoV-2 genomes.  These are shown in Table \ref{tab:seqlist}, along with the RaTG13 and SARS-CoV sequences for this segment.  Further, the BLAST output shows that there are otherwise no prior sequences in the entire database, besides these four, that share more than 10 amino acids in common with this segment of SARS-CoV-2.  Thus, we may emphasize that the 3-D structure of this segment would be especially challenging to predict using computational methods, e.g., as seen in the Robetta results (Section \ref{sec:seq2struct}), as the known structures of existing sequences in the PDB can provide very limited guidance in this case.

\begin{table}[!htbp]
	\caption{Amino acid sequences for positions 469 to 487 in the spike proteins of SARS-CoV-2 and its known mutants, RaTG13, and SARS-CoV.  The Identifier column shows the sequence ID as given in the \texttt{nr} database searched via BLAST; `PDB' indicates that a 3-D structure is available for that protein.  The bold letters in the Sequence column indicate amino acid differences compared to the first known SARS-CoV-2 sequence.}
	\begin{center}
		\begin{tabular}{c|c|c|c|}
			\# & Name &  Identifier & Sequence \\
			\hline
			1 & SARS-CoV-2 original & PDB:6VSB\_A & \texttt{STEIYQAGSTPCNGVEGFN} \\
			2 & SARS-CoV-2 mutant 1 & QIQ49882.1  & \texttt{STEIYQA\textbf{S}STPCNGVEGFN} \\
			3 & SARS-CoV-2 mutant 2 &  QIS30165.1 & \texttt{STEIYQAGSTPCNG\textbf{A}EGFN} \\
			4 & Bat coronavirus RaTG13 &  QHR63300.2 & \texttt{STEIYQAGS\textbf{K}PCNG\textbf{QT}G\textbf{L}N} \\		
			5 & SARS-CoV & PDB:5X58\_A  & \texttt{S\textbf{NVPFSPDGK}PC\textbf{TP-PAL}N} \\
		\end{tabular}
	\end{center}
	\label{tab:seqlist}
\end{table}

For these challenging segments, sampling-based methods can be used to more effectively explore the space of plausible 3-D structures.  Thus in what follows, we assess and quantify the structural uncertainties associated with mutations of this segment, assuming the rest of the SARS-CoV-2 S protein is held fixed.  To do so, we will use an energy function to guide the sampling, for each of the five sequences identified in Table \ref{tab:seqlist}.  A powerful method for this purpose is via sequential Monte Carlo, which we motivate and review next.

\section{Sequential Monte Carlo method for sampling protein structures}

The energy landscape theory of \citet{onuchic1997theory} suggests that a protein structure stabilizes at the 3-D arrangement of atoms, or \textit{conformation}, with the lowest potential energy.  This principle can be leveraged by sampling methods designed to explore the space of possible low energy conformations for a given amino acid sequence.  Suppose $H$ is a given energy function, that takes a conformation $x$ as input, and outputs a scalar value for energy.  Then for a given amino acid sequence, we may conceptualize sampling conformations from the Boltzmann distribution:
\begin{eqnarray}
\pi(x) \propto \exp \left\{-H(x)/T\right\},  \label{pix}
\end{eqnarray}
where $T$ is the effective temperature that may be set to 1 by appropriately scaling the energy function.

In practice, nature’s ‘true’ energy function is not known, and so various approximate energy functions have been developed, often with parameters trained on structure data  such that realistic conformations (i.e., those closer to the `truth' as determined by laboratory techniques) generally have lower energies than those that are not \citep{zhang2007monte}.  All trained energy functions are imperfect, in the sense that a conformation with larger RMSD from the truth may sometimes have lower energy than a conformation with smaller RMSD.  For this study, we adopt an energy function that has been used with reasonable success in \citet{wong2018exploring}, and our previous work shows that it is useful for the purpose of quantifying the space of low-energy conformations.

Protein geometric contraints (including bond lengths and angles) allow $x$ to be more simply parametrized in terms of free dihedral angles, rather than Cartesian coordinates.  Each amino acid in the sequence has three such angles ($\phi, \psi, \omega$) that determine the placements of its backbone atoms, along with 0-4 additional angles denoted $\chi$ governing the placement of its side chain atoms.
The goal then is to draw samples (conformations) from the high-density regions of $\pi(x)$, or equivalently from the low-energy regions.  Considering the five amino acid sequences identified in Table \ref{tab:seqlist}, we note that the sampling problem is difficult:  the segment of interest is 19 amino acids long, which is a high dimensional space with $>60$ geometric degrees of freedom (backbone plus side chain for each amino acid). Further, the energy landscape is highly multimodal and rough due to the numerous pair-wise atomic interactions within the protein.  Due to the difficulty of this sampling problem, most previous sampling methods have focused on shorter segments, e.g., of lengths 12 to 17 \citep{tang2014fast}.

A powerful approach that can be leveraged for this sampling problem comes from Monte Carlo methods, and specifically sequential Monte Carlo (SMC).  The original conformation sampling algorithm for protein segments based on sequential sampling techniques was proposed in \citet{zhang2007biopolymer}, which outperformed other approaches on 2-D and 3-D lattice models.  In subsequent work, sampling methods inspired by SMC were also shown to be successful on real protein structures \citep{tang2014fast,wong2018exploring}. This paper adopts the SMC methodology presented in \citet{wong2018exploring}.  A brief overview of the method is provided here and the interested reader may refer to that paper for details.  The key idea is to sequentially sample the angles of $x$ one amino acid at a time, and this provides a natural incremental distribution for SMC where each particle is a partially constructed conformation.  At the $i$-th SMC step, we sample from the conditional distribution of amino acid $i$ in the sequence, given the previously sampled amino acids, $1,2, \ldots, i-1$. That is, we sample $(\phi_i, \psi_i, \omega_i, \chi_i)$, conditional on $(\phi_{1:i-1}, \psi_{1:i-1}, \omega_{1:i-1}, \chi_{1:i-1})$ when $i>1$, according to an incremental energy function $\Delta H( \phi_i, \psi_i, \omega_i, \chi_i | \phi_{1:i-1}, \psi_{1:i-1}, \omega_{1:i-1}, \chi_{1:i-1})$.  The incremental energy helps to (a) rule out the possibility of atomic clashes, and (b) ensure that the segment connects properly with the two fixed ends of the rest of the protein.  The pool of partial conformations is expanded and then filtered at each step to ensure that a diverse set of low-energy particles is maintained.  A secondary filtering step is embedded to handle the sampling of the amino acid side chains.  The final output of the SMC sampler is a set of sampled conformations for the input sequence, that can be considered to represent the low-energy space of (\ref{pix}).  Thus analysis of these SMC samples can yield insight into the differences between the conformational spaces corresponding to the different input sequences.

We close this section with some technical details.  To obtain our main results, we will apply the SMC method to each of the five amino acid sequences identified in Table \ref{tab:seqlist}.  Since the focus is on potential local structural impacts of amino acid substitutions, we hold the rest of the S protein fixed at the coordinates in the PDB structure 6VSB.  We note that this is a simplifying assumption, as we cannot rule out the possibility that local substitutions can have more global impacts on the protein structure.  Nonetheless, this is common practice in the bioinformatics literature on loop modeling \citep[e.g.,][]{soto2008loop,tang2014fast,wong2017fast,marks2017sphinx}, and likewise we expect this approach to yield useful information here.  Following, we then examine the sampled conformations for the five sequences, and the comparison of their distributions can provide insight into the impacts of the amino acid substitutions. To prepare the data, we take the coordinates from the PDB file 6VSB as the base 3-D structure for the SARS-CoV-2 S protein.  Recall that the 6VSB PDB structure is missing the coordinates of many amino acids in the RBD, while 6LZG is overall complete. By superimposing the coordinates of the 150 RBD amino acids present in both 6LZG and 6VSB, we find a RMSD of 0.95.  Thus, given their overall similarity, we proceed with an approximation by imputing the missing RBD coordinates from 6VSB with those from 6LZG after optimal superposition.  Using this base structure, we then create mutants by substituting the sequences in Table \ref{tab:seqlist} for the segment 469 to 487.  We then use the SMC method to generate conformations for this segment, in all of the sequence variants as well as the base structure, as a basis for comparisons.

\section{Results}

We present our main results in Section \ref{SMCres}:  the SMC method is applied to the sequences listed in Table \ref{tab:seqlist}, and the potential differences in their 3-D conformational spaces are quantified.  Following in Section \ref{sec:bench}, the protein segment prediction accuracies of state-of-the-art methods are evaluated on the known structures of SARS-CoV and SARS-CoV-2, highlighting the need for further advances in structure prediction algorithms.

\subsection{Conformational analysis of SARS-CoV-2 mutant sequences}\label{SMCres}

For each of the five sequences in Table \ref{tab:seqlist}, the SMC method was run multiple (six) times, each with 60000 particles to ensure good coverage of the low-energy conformational space.  This required a total runtime of approximately 3 hours per sequence, on an 8-core Xeon 3.7GHz CPU.  In the subsequent comparative analysis, we kept the 20000 lowest energy conformations among the SMC samples as the representatives for each sequence.  We shall denote these samples as $x^{(k)}_i$, for sequences $i \in \{1,2,3,4,5\}$ and conformations $k \in \{1,2, \ldots, 20000\}$.

To gain insight into the distributions of these conformations in 3-D space, we performed RMSD calculations on the segment of interest, namely positions 469 to 487.  Specifically, we computed pairwise RMSDs between conformations, where we define sets of RMSDs grouped according to sequences $i,j \in \{1,2,3,4,5\}$:
\begin{eqnarray}
d_{ij} &\doteq&  \left\{ \text{RMSD}(x^{(k)}_i, x^{(l)}_j) \right\}  \text{ for all } k,l \in \{1,2, \ldots, 20000\} \text{ if } i \ne j, \nonumber \\ 
\text{and~~~} d_{ii} &\doteq&  \left\{ \text{RMSD}(x^{(k)}_i, x^{(l)}_i) \right\}  \text{ for  } k,l \in \{1,2, \ldots, 20000\} \text{ such that } k \ne l.
\end{eqnarray}
Thus the set $d_{ij}$ approximately represents the distribution that would be obtained by repeatedly drawing one random conformation from the low-energy space of sequence $i$, one random conformation from the low-energy space of sequence $j$, and computing the RMSD between those conformations.  Histograms of $d_{ij}$ then provide a simple way to compare these distributions among the five sequences: if the distribution of $d_{ii}$ is very different from that of $d_{ij}$ for $j \ne i$, then that suggests that the plausible low-energy conformations for the two sequences are located in very distinct regions of 3-D space.  Equivalently, that suggests the amino acid differences between the two sequences can potentially lead to significant changes in the corresponding 3-D structure, including its binding capacity.

\begin{figure}[!htbp]
	\centering
	\includegraphics[scale=0.55]{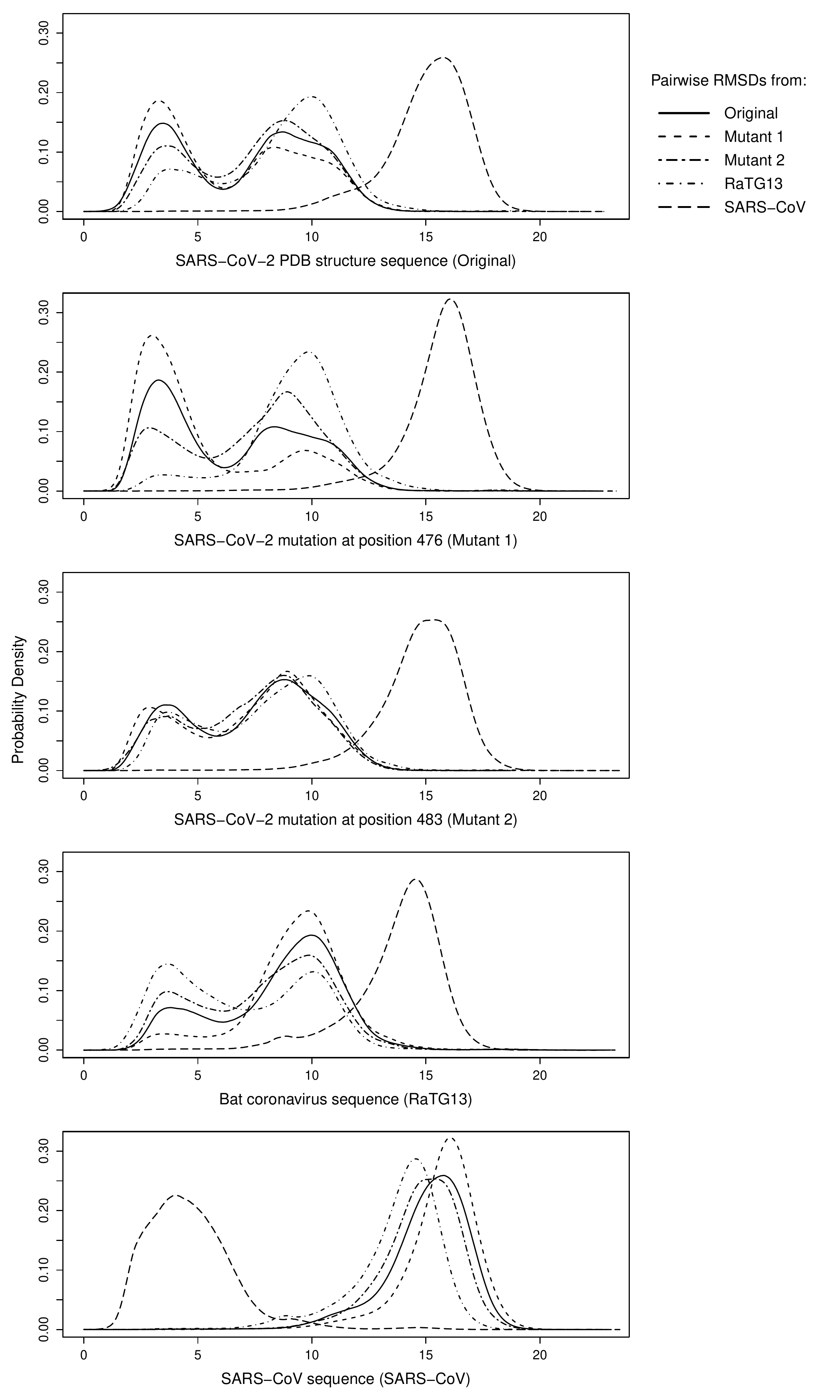}
	\caption{
		Probability densities of the pairwise RMSDs $d_{ij}$, comparing the five sequences in Table \ref{tab:seqlist}.  
		For example, in the top panel, the solid curve shows the distribution of pairwise RMSDs among the 20000 low-energy conformations sampled for the original SARS-CoV-2 sequence, while the long-dashed curve in the same panel shows the distribution of pairwise RMSDs when comparing the samples of the original SARS-CoV-2 sequence with those of SARS-CoV.}
	\label{fig:pairrmsd}
\end{figure}

We plot $d_{ij}$ for all pairs $i,j$ in Figure \ref{fig:pairrmsd}, where the $i$-th panel includes all the distributions $d_{ij}$, $j=1,2,3,4,5$ to facilitate visual comparison.   Kernel density estimation (KDE) was applied to obtain the probability densities shown, using the KDE implementation from \cite{botev2010kernel}.  We may draw several key observations from these results:
\begin{itemize}
\item First, the SARS-CoV amino acid sequence is the most dissimilar from the others, and the bottom panel shows that its low-energy conformational space is also very distinct from the others:  the long-dashed curve shows that most conformations sampled for SARS-CoV are within 2 to 10 RMSD of each other, but mostly $>10$ RMSD away from conformations sampled for the other four sequences.  This relates to Figure \ref{fig:rbdrmsd}: the segment 469 to 487 had large structural differences between the PDB structures of SARS-CoV and SARS-CoV-2, likely due to the large number of amino acid differences in that segment as well as other parts of the sequence.  Our analysis shows that if only this segment of SARS-CoV was substituted into SARS-CoV-2, keeping the rest of the sequence fixed, there would likely still be a large difference between the 3-D structures within that segment.
\item Second, the bat coronavirus RaTG13 sequence appears to have a conformational space that is much more similar to SARS-CoV-2 compared to SARS-CoV, as seen in the fourth panel.  With four amino acid substitutions in the segment 469 to 487 compared to the three SARS-CoV-2 variants, its resulting RMSD distribution could be somewhat distinct from SARS-CoV-2:  RaTG13 has a larger mode at $\sim 3$ RMSD and a smaller mode at $\sim 10$ RMSD compared to all three SARS-CoV-2 variants.
\item Third, the SARS-CoV-2 mutation at position 483 (Mutant 2) appears to possibly have a more substantive effect compared than the mutation at position 478 (Mutant 1).  Mutant 2 involves the substitution of Valine (V) with Alanine (G), which has a less bulky side chain with two fewer methyl (CH$_3$) groups.  The resulting RMSD distributions comparing this mutant to the other SARS-CoV-2 variants and RaTG13 (third panel) become nearly indistinguishable.  Any biological basis for this phenomenon would, of course, require further investigation.
\end{itemize}

\begin{figure}[hp]
	\centering
	{\renewcommand{\arraystretch}{1.04}%
		\begin{tabular}{m{5.5cm}m{4.5cm}}
			\begin{tabular}{c} \textbf{SARS-CoV-2 original} \\ \texttt{STEIYQAGSTPCNGVEGFN}  \end{tabular} &  \includegraphics[width=4cm, trim=0 0 0 0]{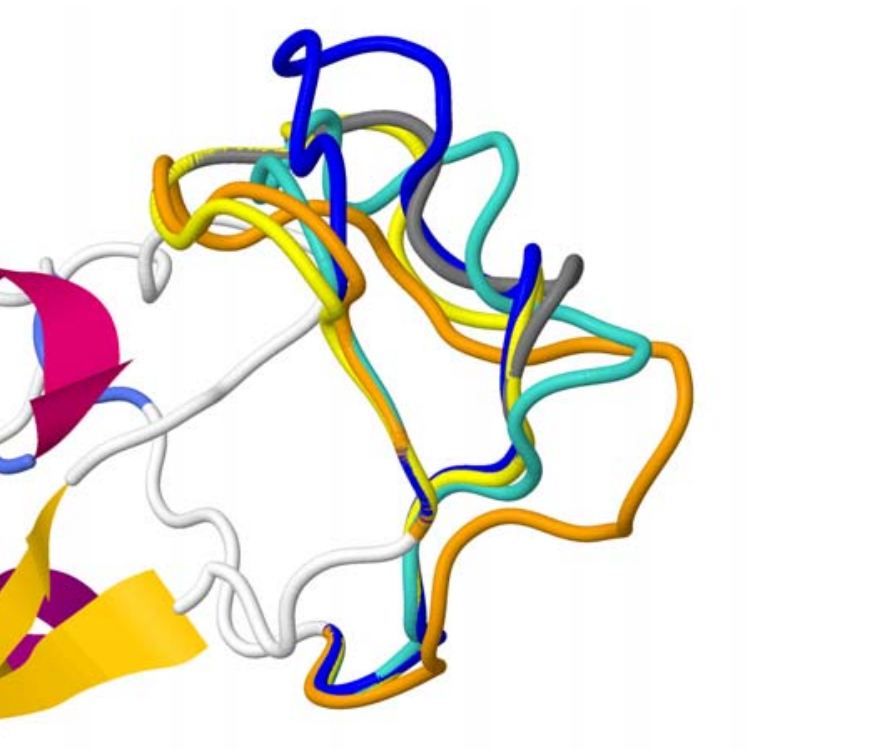}\\ \hline
			\begin{tabular}{c} \textbf{SARS-CoV-2 mutant 1} \\ \texttt{STEIYQA\textbf{S}STPCNGVEGFN}  \end{tabular} & \includegraphics[width=4cm]{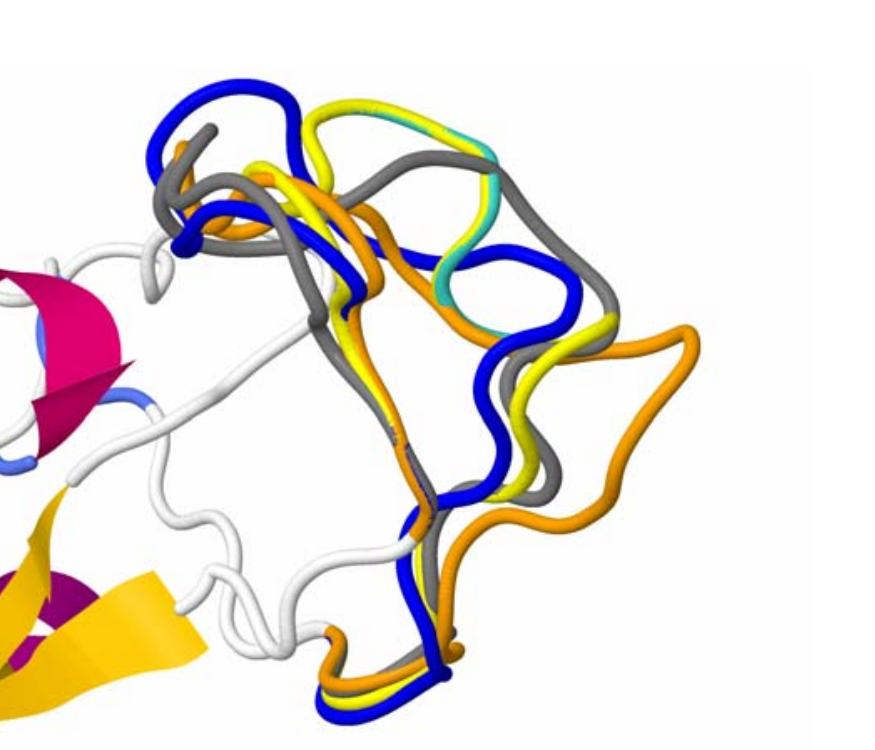}\\ \hline
			\begin{tabular}{c} \textbf{SARS-CoV-2 mutant 2} \\ \texttt{STEIYQAGSTPCNG\textbf{A}EGFN} \end{tabular} & \includegraphics[width=4cm]{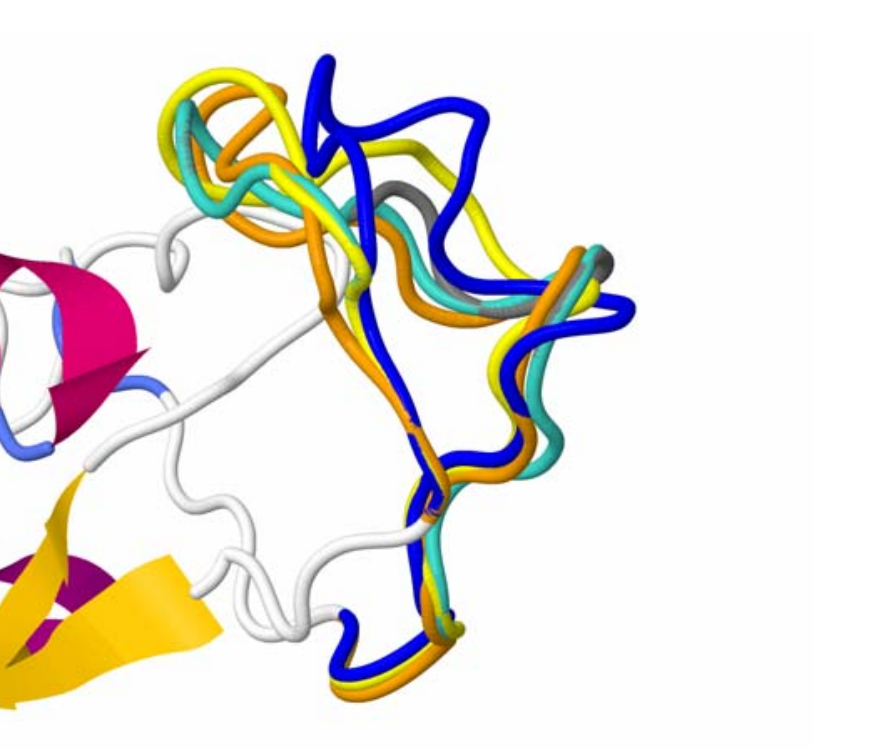}\\ \hline
			\begin{tabular}{c} \textbf{Bat coronavirus RaTG13} \\ \texttt{STEIYQAGS\textbf{K}PCNG\textbf{QT}G\textbf{L}N}  \end{tabular}  & \includegraphics[width=4cm,trim=0 0 30 0,clip]{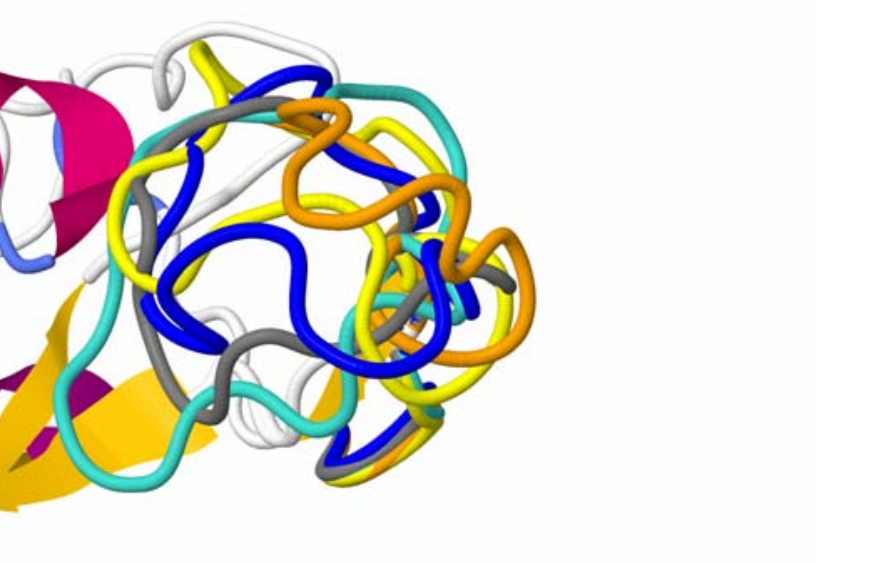} \\ \hline
			\begin{tabular}{c} \textbf{SARS-CoV} \\ \texttt{S\textbf{NVPFSPDGK}PC\textbf{TP-PAL}N}\end{tabular}  & \includegraphics[width=4cm,trim=0 0 50 0]{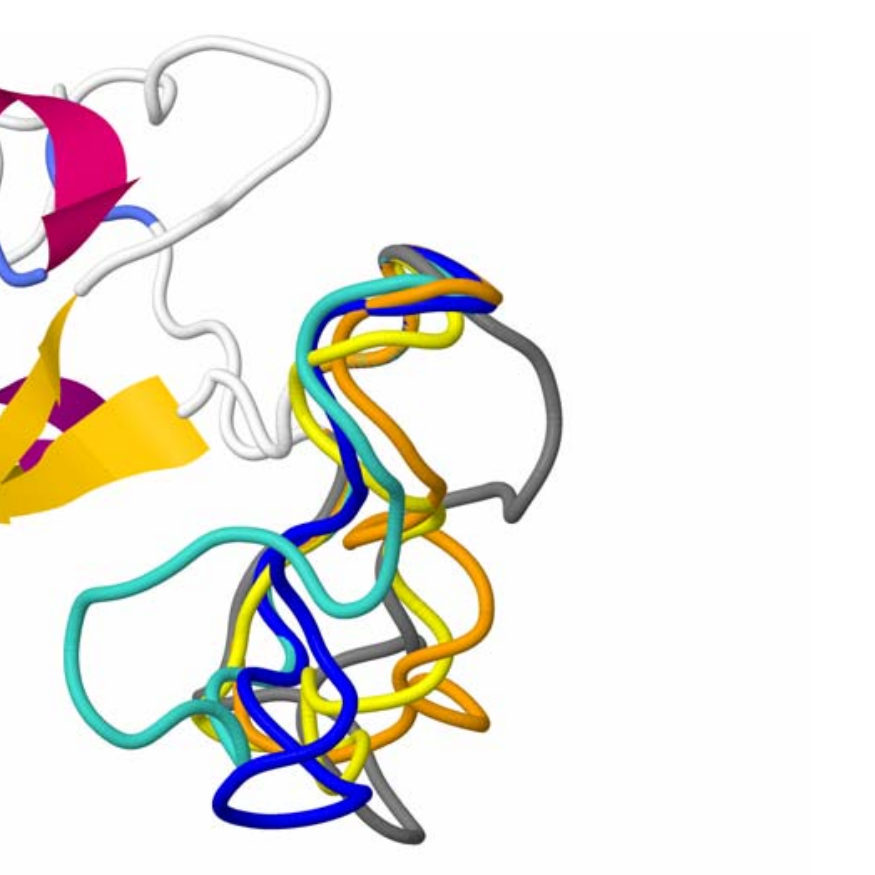}\\
		\end{tabular}
	}
	\caption{For each sequence in Table \ref{tab:seqlist}, the conformations of the five SMC samples with the lowest energy are displayed in these close-ups (colored in order of ascending energy:  grey, turquoise, yellow, blue, orange).  A portion of the fixed part of the protein is visible: the white strands where this segment connects to the rest of the protein, along with some nearby helices and sheets.}
	\label{fig:best5}
\end{figure}

To close this section, in Figure \ref{fig:best5} we visualize the five lowest energy conformations among the SMC samples, for each of the sequences.  Since the segment from 469 to 487 is sampled with the rest of the protein held fixed, the panels show close-ups focusing on the protein backbone for that region of the 3-D structure, along with the nearby amino acids to which this segment is connected.  The five lowest energy conformations are shown with different color strands, as indicated in the figure.  Some structural variability can be observed within these low-energy samples for each sequence, indicating the SMC sampler is successful at discovering distinct conformations with similarly low energy values.  Much greater structural variability can be seen between SARS-CoV-2, RaTG13, and SARS-CoV overall than between the three SARS-CoV-2 variants.

\subsection{Comparison of structure prediction accuracy} \label{sec:bench}

The segment of the RBD studied in this paper poses a challenging test for structure prediction algorithms designed to tackle protein segments.  We evaluate and compare three such recent methods: the SMC method of \citet{wong2018exploring}; the DiSGro method of \citet{tang2014fast}; the next-generation KIC (NGK) method in the Rosetta suite  \citep{stein2013improvements}.\footnote{The Robetta server described in Section \ref{sec:seq2struct} is the automated structure prediction pipeline for entire proteins, built on the Rosetta modeling suite.}  All of these methods are sampling-based and do not rely on the knowledge of existing sequences in the PDB.  We use the known structures of SARS-CoV-2 and SARS-CoV to provide two specific test cases:
\begin{itemize}
	\item Segment 469 to 487 of the SARS-CoV-2 RBD, the main segment of interest in this paper.  We use the true structure provided by PDB code 6LZG, as it has a complete set of atomic coordinates for this segment.
	\item Segment 456 to 473 of the SARS-CoV RBD, which is the corresponding segment in SARS-CoV as shown in Figure \ref{fig:align} in the red box.  The true structure with complete atomic coordinates for this segment is provided by PDB code 5X5B.
\end{itemize}

To prepare a test case, the segment is deleted from the protein while holding the rest of the 3-D structure fixed at the truth.  To assess a method, we use it to draw 500 samples representing the conformational space of the given segment, and the lowest-energy conformation is the method's structure prediction for the missing segment.  Accuracy is evaluated using the RMSD of the reconstructed segments to the known truth.  For the SMC method \citep{wong2018exploring}, we used 60000 particles as in Section \ref{SMCres}, and outputted 500 final conformations.  For the DiSGro method \citep{tang2014fast}, we used the authors' program and increased the default setting of 5000 generated conformations to 100000 to obtain the best possible results; the lowest-energy 500 conformations were kept as the representative samples. For the NGK method \citep{stein2013improvements}, we used the version included in the most recent release of Rosetta 3.12 on April 9, 2020 along with the optimal settings as recommended by the online guide\footnote{https://guybrush.ucsf.edu/benchmarks/benchmarks/loop\_modeling}, and ran the program 500 times to obtain the desired samples.

We note that the DiSGro and SMC methods can each complete the entire sampling and prediction task for a test case in under 30 minutes on an 8-core workstation, with DiSGro requiring only about 10 CPU minutes.  However, NGK is significantly more computationally intensive: generating one sample requires about 45 CPU minutes, so $\sim$2 days is needed to obtain 500 samples on the same 8-core workstation.  The results are shown in Table \ref{tab:methodeval}, where we present two RMSD calculations for each method and test case:  (A) the RMSD of the conformation closest to the truth among the 500 samples, (B) the RMSD of the prediction (i.e., the lowest-energy conformation of the 500 according to that method).  Column (A) shows that in both test cases NGK's samples have a conformation closer to the truth than SMC and DiSGro, while SMC has better samples than DiSGro.   Column (B) shows that the final predictions are not very accurate for any of the methods:  the RMSDs are much larger than column (A), showing that none of the energy functions used by the methods can correctly identify the best RMSD conformation among the samples in this challenging prediction test.  SMC has the best prediction for SARS-CoV-2 (RMSD = 6.97), while NGK has the best prediction for SARS-CoV (RMSD = 4.36).  In particular, the high computational cost of NGK has some apparent advantages for sampling conformations, but has only comparable final prediction accuracy to SMC in these two cases.  These results show that there is an overall need to further improve structure prediction algorithms in future research.

\begin{table}[htbp]
	\centering
	\caption{Evaluating the SMC, DiSGro, and NGK methods on sampling and predicting segments in the known structures of SARS-CoV-2 and SARS-CoV.  The smallest RMSD among the 500 samples and RMSD of the prediction are shown for each method.}
	\centering
	\begin{tabular}{c|ccc|ccc}
		      & \multicolumn{3}{c|}{\textbf{A. Smallest RMSD sampled}} & \multicolumn{3}{c}{\textbf{B. RMSD of prediction}} \\
		 & SMC & DiSGro & NGK   & SMC & DiSGro & NGK \\ \hline
		SARS-CoV-2 & 2.55 & 2.64 & 1.90 & 6.97 & 10.06 &  8.26  \\
		SARS-CoV & 3.24 & 4.02 & 2.18 & 5.02 & 9.60 & 4.36   \\
	\end{tabular}%
	\label{tab:methodeval}%
\end{table}

\section{Discussion and conclusions}

In this paper, we used sequential Monte Carlo to investigate the possible effects to the 3-D structure of the SARS-CoV-2 spike protein due to mutations in its amino acid sequence.  SMC is potentially a powerful technique for this purpose, as it is effective at sampling conformations with low energy for the segment of interest.  Thus, by comparing the sampled conformations for several sequences, the method can help detect changes to the low-energy regions of the 3-D conformational space as a result of the amino acid differences among the sequences.  Our results are consistent with the observed differences between the known 3-D structures for the specific variants of SARS-CoV-2 and SARS-CoV in the Protein Data Bank, and also provide some preliminary intuition about the potential role of mutations that are being observed in SARS-CoV-2.  Given the current public health concern posed by COVID-19 as a result of the SARS-CoV-2 virus, potential effects of mutations to key viral proteins are of high importance, as structural changes may affect the efficacy of ongoing medical developments.  Thus, our work here highlights one problem where sampling methods and data scientists can have an important role.

We view this paper as a foray into the area that combines data science with structural biology for COVID-19 research, that we hope can spur further statistical and scientific investigations.  Thus we list some limitations of the current study along with potential directions to take.  First, there may be other interesting ways to summarize and compare the SMC samples from different sequences.  Here, we extracted the 20000 conformations with the lowest energies and treated them equally; for example, all the samples might be kept and  weighted according to energies for a more comprehensive analysis.  Second, here we only investigated one segment of the SARS-CoV-2 spike protein, amino acids 469 to 487.  If mutations occur in different locations of the sequence, it could be useful to simultaneously sample conformations for multiple disjoint segments together, e.g., \citet{tang2015conformational}.  Third, we took the energy function as given as in \citet{wong2018exploring}.  Different energy functions could be tried to provide further insights.  Fourth, we note that this work is inherently exploratory in nature:  insights about a protein's conformational space discovered in this way would require solid scientific corroboration to have a solid biological basis, given the need for further advances in the accuracy of structure prediction algorithms in general.  Thus our work is consistent with the role of computational methods for protein folding that was noted in the Introduction: computation can complement and help provide focus to ongoing medical and scientific efforts. 

\section*{Acknowledgements}
This work was partially supported by a Discovery Grant from the Natural Sciences and Engineering Research Council of Canada.

\bibliographystyle{apalike}
\bibliography{covid19-protein}

\end{document}